\documentclass[sigconf,nonacm]{acmart}
\renewcommand\footnotetextcopyrightpermission[1]{}
\usepackage{booktabs}
\usepackage{tabularx}
\usepackage{array}
\usepackage{multirow}
\usepackage{amsmath}
\usepackage{xspace}
\usepackage{tikz}
\usetikzlibrary{arrows.meta,positioning,calc,fit}
\usepackage{balance}
\usepackage{graphicx}
\graphicspath{{./}{figures/}}

\AtBeginDocument{}

\newcommand{\ind}{\mathbf{1}}
\newcommand{\E}{\mathbb{E}}
\newcommand{\Q}{\mathbb{Q}}
\newcommand{\R}{\mathbb{R}}
\newcommand{\calD}{\mathcal{D}}
\newcommand{\calH}{\mathcal{H}}
\newcommand{\NoEvent}{\textsc{No-Event Poly}\xspace}
\newcommand{\EGauss}{\textsc{Event Gaussian}\xspace}
\newcommand{\EMix}{\textsc{Event Mixture}\xspace}
\newcommand{\NoEventContam}{\textsc{Contaminated}\xspace}
\newcommand{\NoEventSSVI}{\textsc{No-Event SSVI}\xspace}
\newcommand{\EGaussSSVI}{\textsc{Event Gaussian SSVI}\xspace}
\newcommand{\ENeural}{\textsc{Event Neural-MDN}\xspace}
\newcommand{\ENeuralCal}{\textsc{Event Neural-MDN-Calibrated}\xspace}
\begin{document}
\title{Non-Spanning Identification of Scheduled Event Risk in Option Pricing}
\author{Tenghan Zhong}
\email{tenghanz@usc.edu}
\affiliation{%
  \institution{University of Southern California}
  \city{Los Angeles}
  \state{CA}
  \country{USA\texorpdfstring{\strut}{}}
}
\renewcommand{\shortauthors}{Tenghan Zhong}

\begin{abstract}
Short-dated index options make scheduled macro-announcement risk visible in market prices, but visibility does not imply identification: a flexible no-event surface fitted to event-spanning quotes can absorb event premia, while a jump calibrated without event-spanning quotes is unidentified. To separate the continuous surface from the scheduled jump, we model Federal Open Market Committee (FOMC) decisions, Consumer Price Index (CPI) releases, and nonfarm payroll (NFP) reports as deterministic-time jumps in risk-neutral option pricing and propose a non-spanning identification protocol. Non-spanning expiries identify the no-event volatility surface, event-spanning training quotes calibrate the scheduled jump, and held-out event-spanning quotes are used only for pricing evaluation. On PM-settled S\&P 500 index (SPX) options from May 2022 to August 2025, Gaussian and two-component mixture jumps improve held-out event-spanning pricing, with the clearest gains in robust median pricing errors and in event-volatility option combinations (straddles and strangles) rather than directional risk reversals. A contaminated-surface stress test confirms the identification concern: allowing event-spanning training quotes into the no-event surface fit produces strong held-out performance by absorbing event premia rather than identifying scheduled jump risk. An amortized mixture density network (MDN) benchmark shows limited cross-event transfer: pure leave-one-event-out amortization reduces implied-volatility errors but not mean dollar or mean spread-normalized pricing errors, while the scale-calibrated variant restores Gaussian-level performance yet remains below event-specific mixture calibration. Scheduled-jump identification is strongest for CPI and FOMC and weaker for NFP.
\end{abstract}

\keywords{option pricing, scheduled event risk, non-spanning identification, machine learning in finance, mixture density networks, amortized calibration, short-dated options, implied volatility}

\maketitle

\section{Introduction}
\label{sec:intro}
Short-dated index options reveal how markets price scheduled macroeconomic announcements \cite{Wright2020,LondonoSamadi2023,FedFEDS2025}. When an option expires immediately after a Federal Open Market Committee (FOMC) decision, Consumer Price Index (CPI) release, or nonfarm payroll (NFP) employment report, its price includes a discrete announcement component absent from otherwise similar contracts maturing before the announcement \cite{DubinskyJohannes2004,Wright2020,LondonoSamadi2023}. This visibility creates an identification problem: because standard implied-volatility functions and smoothed surfaces are flexible enough to fit rich strike-maturity panels \cite{DumasFlemingWhaley1998,Fengler2009,GatheralJacquier2014}, a surface fitted to event-spanning contracts can absorb scheduled event premia without explicitly representing the event.

The existing literature mainly uses options to measure announcement uncertainty, event premia, and which calendar days matter for markets \cite{Wright2020,LondonoSamadi2023,FedFEDS2025,KnoxEtAl2026}. Related event-risk studies also show that announcements can change option-surface shape and induce non-Gaussian or bimodal risk-neutral distributions \cite{DubinskyJohannes2004,Alexiou2025}. This paper asks two linked questions. First, does an explicit scheduled jump improve held-out event-spanning prices when the continuous surface is identified only from non-spanning expiries? Second, does the shape of the event-jump density transfer across announcements, or must the target event still provide its own jump scale?

The identification problem follows directly from the maturity structure of a known-date jump. In a deterministic-time-jump model, the scheduled jump enters only contracts whose maturities cross the announcement. Non-spanning expiries can identify the no-event surface within the chosen surface class, but they contain no direct information about the jump. Event-spanning quotes are necessary for jump calibration. They cannot, however, also define the no-event surface without contaminating the object being identified. We therefore separate the information sets: non-spanning expiries identify the no-event surface, event-spanning training quotes calibrate the scheduled jump, and held-out event-spanning contracts are used only for evaluation. The same information restriction extends to the neural benchmarks: pure amortization tests cross-event jump-shape transfer without target-event jump calibration, while the calibrated variant allows only one target-event scale parameter before held-out evaluation.

Our empirical analysis uses PM-settled S\&P 500 index (SPX) options from May 2022 through August 2025 and focuses on the three announcement types. Held-out tests show that explicit scheduled jumps improve event-spanning pricing relative to a no-event extrapolation, with the strongest evidence for CPI and FOMC. Additional benchmarks clarify the mechanism: the surface stochastic-volatility-inspired (SSVI) check rules out a polynomial-surface artifact, the contaminated-surface stress test exposes the failure mode of letting event-spanning training quotes enter the no-event surface fit, option-combination pricing tests locate the gains in event-volatility combinations, and neural benchmarks show that cross-event jump-shape transfer is limited.

The paper makes three contributions.

\begin{itemize}

\item \textbf{A non-spanning identification protocol for scheduled event risk.}
We propose a protocol that separates no-event surface identification from scheduled-jump calibration by whether a contract's maturity spans the target event, with a stratified subset of event-spanning contracts held out solely for pricing evaluation. Inference is conducted at the event level; same-day joint announcements are removed, and neighboring events are isolated through quote-date and expiry filters.

\item \textbf{Empirical evidence on scheduled event-risk pricing.}
Under the proposed protocol, Gaussian and two-component mixture event-jump models improve held-out event-spanning SPX option pricing. The SSVI surface-family robustness check and the contaminated-surface stress test support the identification interpretation of this gain over surface absorption of event premia, and option-combination pricing tests locate it in event-volatility combinations rather than directional risk reversals.

\item \textbf{Amortized density learning for scheduled-risk transfer.}
We construct a mixture density network (MDN) benchmark to test whether scheduled jump-density shape transfers across announcements under the non-spanning information restriction. Held-out comparisons yield a scale-shape attribution: cross-event learning transfers some jump-shape information, target-event event-spanning training quotes identify most of the jump scale, and event-specific mixture calibration supplies the remaining higher-moment improvement. This identifies an informational boundary for amortized calibration of scheduled event risk rather than an optimization failure.

\end{itemize}

\section{Related Work}
\label{sec:related}

\paragraph{Event-day options and macro announcements.} A growing literature uses short-dated options to measure event-day uncertainty and risk premia. Wright \cite{Wright2020} studies risk-neutral uncertainty and variance risk
premia around FOMC and employment releases, while Londono and Samadi
\cite{LondonoSamadi2023} use daily S\&P 500 expirations to estimate
release-specific premia for CPI, FOMC, NFP, and gross domestic product (GDP).
Building on this measurement agenda, recent work moves from named macro
releases to event-day classification: Federal Reserve research infers priced
equity-market event days from daily-expiration options \cite{FedFEDS2025}, and
Knox et al. \cite{KnoxEtAl2026} identify elevated option-implied equity premium
days, emphasizing FOMC, employment, and CPI releases. We extend this measurement literature from estimating event premia to validating whether an explicit scheduled-jump component improves unseen event-spanning prices after the no-event surface is fixed from non-spanning expiries.

\paragraph{Event-risk option surfaces.} Option prices encode state-price density information \cite{BreedenLitzenberger1978}, so announcement risk can produce option-surface shapes that are not well summarized by a single smooth variance level. Dubinsky and Johannes \cite{DubinskyJohannes2004} model known-date earnings jumps in equity options, and Alexiou et al. \cite{Alexiou2025} show that earnings-announcement risk can generate concavity and bimodality in short-term risk-neutral distributions. We move this surface-shape evidence to index-level macro events and require the event-jump specification to improve unseen event-spanning contracts, including event-volatility option combinations, rather than merely produce a plausible event-day density.

\paragraph{Option pricing, calibration, and model validation.} Classical option pricing starts with no-arbitrage diffusion and stochastic-volatility models \cite{BlackScholes1973,Heston1993} and jump extensions \cite{Merton1976,Bates1996}. Practical calibration often uses deterministic implied-volatility functions or smoothed implied-volatility surfaces \cite{DumasFlemingWhaley1998,GatheralJacquier2014,Fengler2009}, with deep learning, hypernetwork, and neural-operator surface smoothing approaches providing additional flexibility \cite{Ackerer2020,YangChenShuHospedales2025,WiedemannJacquierGonon2025}. Short-maturity SPX option panels are a direct laboratory for volatility and jump risks and for recovering dynamic state variables \cite{AndersenFusariTodorov2017,AndersenFusariTodorov2015,TodorovZhang2022}. Our model is deliberately structured: FOMC, CPI, and employment announcements are deterministic-time jumps, not randomly arriving Poisson shocks. Unlike same-date surface calibration or smoothing exercises, our evaluation withholds event-spanning contracts and asks whether an explicit scheduled jump improves held-out pricing after the continuous surface is fixed by non-spanning expiries.

\paragraph{Neural and amortized calibration.} We use neural methods not as a replacement for the option surface, but as an amortized test of scheduled-jump transferability. Learning-based derivative pricing dates back to neural-network approximations of pricing and hedging rules \cite{HutchinsonLoPoggio1994}. Neural methods now learn maps from model parameters or market features to prices and implied volatilities \cite{Hernandez2017,BayerStemper2018,LiuOosterleeBohte2019}. Related approaches learn calibrated parameters or neural volatility representations \cite{CuchieroKhosrawiTeichmann2020,HorvathMuguruzaTomas2021}. Mixture density networks provide the classical conditional-density architecture behind our neural jump density \cite{Bishop1994}. Recent neural smoothing methods map sparse or irregular option observations directly to arbitrage-aware implied-volatility surfaces and benchmark against stochastic-volatility-inspired (SVI) surfaces \cite{YangChenShuHospedales2025,WiedemannJacquierGonon2025}. These methods are valuable for same-date surface construction, surrogate pricing, and fast recalibration. Our setting differs in two ways. First, the scheduled jump is a low-dimensional addition to the continuous surface rather than the surface itself. Second, the neural density is evaluated under the same non-spanning protocol as the parametric jumps, which isolates cross-event transferability of scheduled jump structure from continuous-surface flexibility.

\section{Identification Protocol and Pricing Models}
\label{sec:model}
For a scheduled event date $e$, the protocol starts from a pre-event quote date
$t$ and partitions each option by whether its expiration crosses $e$; when
indexing multiple events, $t_e<e$ is the selected quote date for event date $e$. Contract
$i$ observed at $t$ has strike $K_i$, expiration $T_i$, bid $b_i$, ask $a_i$,
midpoint $m_i=(a_i+b_i)/2$, and call/put type. Let $F_t(T_i)>0$ and
$D_t(T_i)>0$ denote the forward price and discount factor for expiration $T_i$,
and write $\E^\Q[\cdot]$ for risk-neutral expectation. Define log-moneyness and
time to maturity as $k_i=\log\{K_i/F_t(T_i)\}$ and $\tau_i=(T_i-t)/365$. The
event-spanning indicator is $I_{i,t,e}=\ind\{T_i\ge e\}$. Contracts with
$I_{i,t,e}=0$ expire before the target event and are used to fit the no-event
surface; contracts with $I_{i,t,e}=1$ span it and contain the scheduled event
component. Event-date expirations are event-spanning because all three
announcements occur before PM settlement. Let $\calD^{0}_{t,e}$ denote
non-spanning contracts, $\calD^{1,\mathrm{tr}}_{t,e}$ event-spanning training
contracts, and $\calD^{1,\mathrm{ho}}_{t,e}$ held-out event-spanning contracts.

The key restriction is that the continuous surface is fitted only on
$\calD^{0}_{t,e}$, event jumps are calibrated only with
$\calD^{1,\mathrm{tr}}_{t,e}$, and $\calD^{1,\mathrm{ho}}_{t,e}$ is used only
after all parameters are fixed.

\subsection{Pricing Model: Continuous Surface and Event Jumps}
The no-event model, denoted \NoEvent in the tables, prices contract $i$ with a
contract-indexed Black--Scholes-equivalent continuous return,
\begin{equation}
X^c_{t,i}\sim \mathcal{N}\left(-\frac{1}{2}w_c(k_i,\tau_i), w_c(k_i,\tau_i)\right).
\end{equation}
Here $X^c_{t,i}$ is a local pricing return for contract $i$, indexed by
$(k_i,\tau_i)$ rather than a common terminal return shared across strikes. The continuous total variance is parameterized through a bounded variance-rate surface,
\begin{equation}
w_c(k,\tau)=\tau\sigma_c^2(k,\tau),\qquad
\log\sigma_c^2(k,\tau)=b(k,\tau)^\top\beta,
\end{equation}
where $b(k,\tau)\in\R^p$ is a fixed polynomial basis, $p$ is the basis dimension, and $\beta\in\R^p$ is the fitted coefficient vector. Let $\widehat w_c$ denote
the fitted continuous surface obtained from $\calD^0_{t,e}$. The implied continuous volatility is clipped to $[0.03,c_{\max}]$ ($c_{\max}=1.00$ in SPX).
Because $w_c$ is fitted as a strike-maturity surface, we interpret it as a
$(k,\tau)$-indexed Black--Scholes-equivalent pricing control rather than a
joint risk-neutral law across expirations. The event jump is added separately
as a strike-independent, martingale-normalized component, so identification is
conditional on the chosen no-event surface class.

Since announcement times are known at the quote date, we use deterministic-time
jumps instead of random-arrival Poisson jumps \cite{Merton1976,Bates1996}: a
maturity either crosses the scheduled date or it does not. For any contract in the pricing panel, the model-implied log return is
\begin{equation}
R_{\theta,i}=X^c_{t,i}+J_{e,t}I_{i,t,e},
\label{eq:event_return}
\end{equation}
where $R_{\theta,i}$ is the model-implied log return for contract $i$, $\theta$
collects the active parameters of the current specification, and $X^c_{t,i}$
and $J_{e,t}$ are independent under $\Q$. \EGauss uses
\begin{equation}
J_{e,t}\sim \mathcal{N}\left(-\frac{1}{2}s_e^2,s_e^2\right),
\end{equation}
where $s_e>0$ is the event-jump standard deviation, so that $\E^\Q[\exp(J_{e,t})]=1$. \EMix uses an unnormalized two-component mixture
\begin{equation}
\widetilde J_{e,t}\sim\sum_{m=1}^2 p_m\mathcal{N}(\mu_m,\sigma_m^2),
\end{equation}
where $p_m\ge0$, $\sum_{m=1}^2 p_m=1$, $\mu_m\in\mathbb{R}$, and $\sigma_m>0$. It then applies the martingale normalization
\begin{equation}
A_e=\log\left(\sum_{m=1}^2p_m\exp\{\mu_m+\sigma_m^2/2\}\right),
\qquad
J_{e,t}=\widetilde J_{e,t}-A_e,
\end{equation}
which gives $\E^\Q[\exp(J_{e,t})]=1$. 

\begin{figure*}[htp]
\centering
\includegraphics[width=\textwidth]{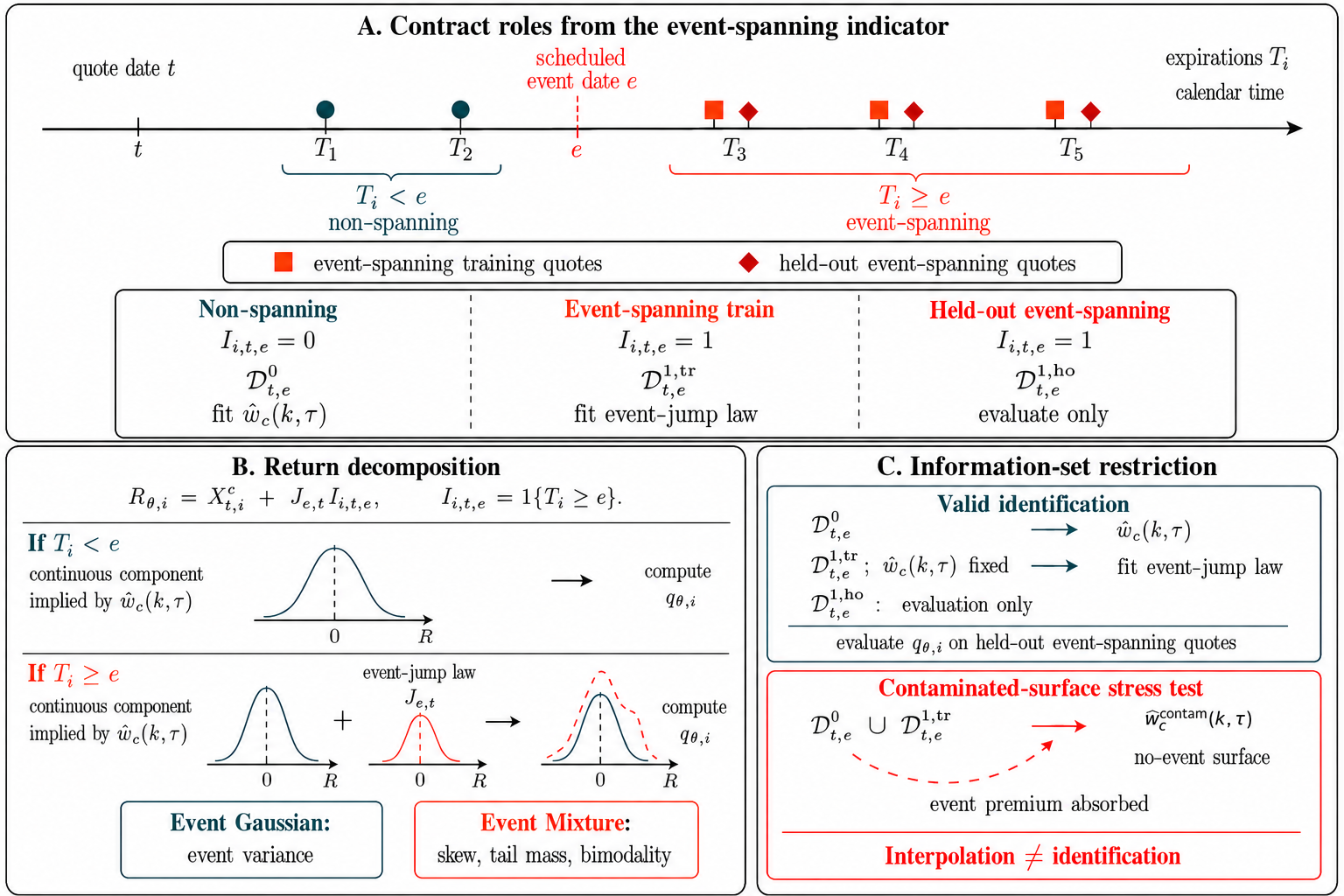}
\caption{
Non-spanning identification for scheduled event risk.
Contracts with $T_i<e$ identify the continuous Black--Scholes-equivalent surface
$\widehat w_c(k,\tau)$.
Event-spanning training quotes fit the event-jump law for $J_{e,t}$, while held-out event-spanning quotes evaluate pricing
only after all parameters are fixed.
The contaminated-surface stress test shows how event-spanning training quotes
can contaminate the continuous surface by absorbing event premia.
}
\Description{Three-panel mechanism diagram. Panel A shows quote date, scheduled event date, non-spanning expiries, event-spanning training quotes, and held-out event-spanning quotes. Panel B shows the return decomposition into continuous and scheduled event-jump components. Panel C contrasts valid non-spanning identification with the contaminated-surface stress test.}
\label{fig:protocol}
\end{figure*}
\paragraph{Amortized neural event-jump density.}
\ENeural maps a non-spanning pre-event conditioning vector $c_{e,t}$ through a
neural network $g_\phi$, with weights $\phi$, to $M=4$ mixture weights, means,
and standard deviations. Writing $c=c_{e,t}$, the network returns mixture logits $z_m(c)$, mean outputs
$\mu_m(c)$ clipped to $[-0.10,0.10]$, and scale logits $\eta_m(c)$. The mixture
weights and component scales are
$p_m(c)=\exp\{z_m(c)\}/\sum_{j=1}^M\exp\{z_j(c)\}$ and
$\sigma_m(c)=\mathrm{clip}\{10^{-4}+\mathrm{softplus}(\eta_m(c)),10^{-4},0.10\}$. The unnormalized jump is
\[
\widetilde J_{\phi,e,t}\mid c_{e,t}\sim\sum_{m=1}^M p_m(c_{e,t})\,\mathcal{N}\!\left(\mu_m(c_{e,t}),\sigma_m^2(c_{e,t})\right),
\]
martingale-normalized by
\[
J_{\phi,e,t}=\widetilde J_{\phi,e,t}-\log\sum_{m=1}^M p_m(c_{e,t})\exp\!\left\{\mu_m(c_{e,t})+\tfrac{1}{2}\sigma_m^2(c_{e,t})\right\},
\]
so that $\E^\Q[\exp(J_{\phi,e,t})\mid c_{e,t}]=1$. The conditioning vector contains event-type indicators, days to event, non-spanning ATM IV, non-spanning skew and term-slope proxies, and pre-quote realized volatility and return measures. Event-spanning quotes, release values, macro surprises, event-day returns, and held-out quote statistics are excluded. The network is trained leave-one-event-out: for held-out event $e$, $g_\phi$ is fitted on $\bigcup_{e'\neq e}\calD^{1,\mathrm{tr}}_{t_{e'},e'}$ by backpropagating the pricing loss \eqref{eq:loss} through analytic mixture prices, with conditioning constructed only from non-spanning contracts of event $e'$ and pre-quote-date realized state.

\paragraph{Semi-amortized calibrated variant.}
The calibrated variant fixes the leave-one-event-out neural mixture shape and estimates a single target-event scale parameter $\gamma_e\in[0,5]$ on $\calD^{1,\mathrm{tr}}_{t,e}$:
\[
\begin{aligned}
J^{\mathrm{cal}}_{\phi,e,t}
&=\gamma_e\widetilde J_{\phi,e,t}-A_\gamma(c_{e,t}),\\
A_\gamma(c_{e,t})
&=\log\sum_{m=1}^M p_m(c_{e,t})
\exp\!\left\{\gamma_e\mu_m(c_{e,t})+\tfrac{1}{2}\gamma_e^2\sigma_m^2(c_{e,t})\right\}.
\end{aligned}
\]
\ENeuralCal is therefore a semi-amortized scale-calibration benchmark.

For a contract with model-implied log return $R_{\theta,i}$, the model price is
\begin{equation}
q_{\theta,i}=D_t(T_i)
\begin{cases}
\E^\Q\left[(F_t(T_i)e^{R_{\theta,i}}-K_i)^+\right], & \text{call},\\[0.3ex]
\E^\Q\left[(K_i-F_t(T_i)e^{R_{\theta,i}})^+\right], & \text{put}.
\end{cases}
\label{eq:model_price}
\end{equation}
The expectation is analytic for \NoEvent and \EGauss, and a weighted sum of Black--Scholes prices for \EMix, \ENeural, and \ENeuralCal. For \EMix, each mixture component is priced as a Black--Scholes contract with total variance $w_c(k_i,\tau_i)+\sigma_m^2$ and component forward multiplier $\exp\{\mu_m-A_e+\sigma_m^2/2\}$, avoiding simulation noise. Neural variants use the same formula with predicted or $\gamma_e$-scaled mixture parameters.

\paragraph{Jump moments reported for interpretation.}
For \EGauss, the event jump variance is simply $s_e^2$. For \EMix, the normalized mixture distribution implies
$\E^\Q[J_{e,t}]=\sum_{m=1}^2 p_m(\mu_m-A_e)$. The reported jump standard deviation uses the standard mixture variance decomposition. These summaries help interpret the fitted jump and are not used as training targets. Figure~\ref{fig:protocol} gives the corresponding information-set schematic.

\subsection{Calibration Loss and Implementation}
Let $s_i=\max\{a_i-b_i,0.02m_i,10^{-4}\}$ denote the robust spread scale used for calibration. The calibration loss on a training set $\calD$ is
\begin{equation}
\mathcal{L}(\theta;\calD)=
\sum_{i\in\calD}\omega_i\rho\left(\frac{q_{\theta,i}-m_i}{s_i}\right)+\lambda\|\theta\|_2^2,
\label{eq:loss}
\end{equation}
where $\rho$ is the Huber loss with transition $\delta=3.0$, and $\theta$ denotes the parameters fitted in the current calibration stage. The weight $\omega_i$ is vega divided by relative spread $(a_i-b_i)/m_i$, median-normalized, and clipped to $[0.2,5]$; $\lambda\ge0$ is a ridge penalty. The calibration loss uses the robust spread scale $s_i$, while reported spread-normalized evaluation errors use the quoted bid-ask spread floored at $10^{-4}$. Thus, differences across reported specifications come from the risk representation between quote date and expiration, not from data filters, held-out assignment, liquidity weights, or the training loss.

\paragraph{Implementation.}
The continuous basis contains seven terms: constant, $k$, $k^2$, $\tau_{30}$, $\sqrt{\tau_{30}}$, $k\tau_{30}$, and $\tau_{30}^2$, where $\tau_{30}=\tau/(30/365)$. Continuous and jump ridge penalties are $10^{-4}$ and $10^{-3}$, respectively. Gaussian $s_e$ and mixture component standard deviations are bounded to $[0.001,0.50]$. Parametric continuous and jump fits use bound-constrained L-BFGS-B with 500 iterations. \EMix uses ordering-penalty weight $10^{-3}$ on $\max(\mu_1-\mu_2,0)^2$. The SSVI check replaces only the continuous-surface fit with a compact SSVI surface using the power-law shape function of Gatheral and Jacquier \cite{GatheralJacquier2014}, leaving all other protocol steps unchanged. The neural MDN is a two-layer width-32 GELU MLP trained with Adam, learning rate $10^{-3}$, a 10\% event-level validation split, and early stopping; \ENeuralCal fits $\gamma_e$ by bounded scalar minimization. Reported neural predictions average prices across three seeds.

\subsection{Event-Level Estimation and Test Split}
\label{sec:split}
The non-spanning identification protocol can be written as a repeated event-level algorithm. With the contract groups defined above, the main specification applies the following steps to every usable event:
\begin{enumerate}
    \item Fit $\beta$ in the continuous variance-rate surface using $\calD^{0}_{t,e}$.
    \item Price held-out event-spanning contracts with the no-event extrapolation $J_{e,t}=0$.
    \item Holding $\beta$ fixed, fit $s_e$ for \EGauss using $\calD^{1,\mathrm{tr}}_{t,e}$ and price $\calD^{1,\mathrm{ho}}_{t,e}$.
    \item Holding $\beta$ fixed, fit $(p_m,\mu_m,\sigma_m)_{m=1}^2$ for \EMix using $\calD^{1,\mathrm{tr}}_{t,e}$ and price $\calD^{1,\mathrm{ho}}_{t,e}$.
    \item Train $g_\phi$ on $\bigcup_{e'\neq e}\calD^{1,\mathrm{tr}}_{t_{e'},e'}$ for \ENeural with conditioning $c_{e',t_{e'}}$ from non-spanning contracts and pre-event state only. Set the jump density for event $e$ to $g_{\widehat\phi_{-e}}(c_{e,t})$, where $\widehat\phi_{-e}$ denotes neural weights fitted on all events except $e$, and price $\calD^{1,\mathrm{ho}}_{t,e}$ without target-event jump calibration.
    \item For \ENeuralCal, hold $g_{\widehat\phi_{-e}}(c_{e,t})$ fixed and fit $\gamma_e\in[0,5]$ on $\calD^{1,\mathrm{tr}}_{t,e}$ using the same Huber-on-spread-normalized loss as the parametric jumps; then price $\calD^{1,\mathrm{ho}}_{t,e}$.
    \item Evaluate held-out quote errors, event-level averaged paired differences, option-combination errors, and fitted jump summaries.
\end{enumerate}
These steps define the event-level information set and unit of inference. For \ENeural, the target event is also excluded from feature standardization. Held-out quotes are never used for fitting.

Before applying quote-count floors, the split enforces calendar isolation. For target event $e$, the selected quote date must fall after the previous target event in the filtered target-event calendar. Event-spanning candidates must expire before the next target event in the filtered target-event calendar and no later than five calendar days after $e$. Among pre-event trading dates passing these isolation filters, the main specification selects the quote date nearest to the target event subject to the quote-count floors below. An event then enters only if the selected quote date has at least 60 combined training quotes (non-spanning plus event-spanning training), with separate floors of 20 event-spanning training, 10 non-spanning training, and 20 held-out event-spanning quotes. Held-out contracts are sampled only from event-spanning quotes (25\% per event) and stratified by log-moneyness bin and call/put type. The randomization is seeded by event identifier and quote date.

\subsection{Held-Out Metrics and Event-Level Inference}
Given the held-out set produced by this split, we evaluate price MAE,
spread-normalized MAE, median spread-normalized absolute error, bid-ask
containment, IV MAE, and median IV error. Let
$\bar{s}_i=\max\{a_i-b_i,10^{-4}\}$ denote the quoted-spread scale used for
reported evaluation. For a held-out set $\calH$, spread-normalized MAE is
$|\calH|^{-1}\sum_{i\in\calH}|q_{\theta,i}-m_i|/\bar{s}_i$, and bid-ask
containment is
$|\calH|^{-1}\sum_{i\in\calH}\ind\{b_i\le q_{\theta,i}\le a_i\}$.
For IV metrics, model prices are converted to Black--Scholes implied
volatilities using the same forward and discount factor as the market quote;
errors are measured against the market midpoint IV and computed only when the
model price is invertible. Pairwise-common IV comparisons, restricted to
contracts invertible under both models in a pair, give the same sign as the
model-specific IV columns reported in the tables.

Since held-out quotes around the same event are dependent, confidence
intervals use event-level paired bootstrap rather than quote-level resampling.
For a candidate specification $r$ and baseline specification $0$, let
$\ell_{r,i}$ and $\ell_{0,i}$ denote their losses on contract $i$. We first
average the difference $\ell_{r,i}-\ell_{0,i}$ within each event identifier,
then bootstrap those event-level differences with 2,000 resamples and report
percentile intervals.

\section{SPX Data and Evaluation Design}
\label{sec:data}
The main sample uses SPX options from OptionMetrics from 2022-05-01 to
2025-08-29. This period offers daily expirations, so both non-spanning and
event-spanning maturities are available near every announcement. The event
calendar contains FOMC decision dates, CPI release dates, and NFP employment
releases. FOMC events use the meeting end date, and CPI/NFP events use public
release dates. We first remove days on which target events coincide, leaving
105 scheduled event dates. Applying the quote-date selection rules, calendar
isolation filters, and quote-count floors of Section~\ref{sec:split} leaves
101 usable events and 11,135 held-out quotes. The main specification, SSVI
surface-family robustness check, and contaminated-surface stress test use the
same event and held-out panels.

The event calendar is intentionally minimal. We do not use release values or surprises as model inputs. The calendar only determines whether a maturity spans a known scheduled event. This keeps the pricing problem risk-neutral and avoids turning the paper into a macro-surprise forecasting exercise.

The main market is SPX because it is the primary index-option market used in recent macro-announcement option studies \cite{FedFEDS2025,KnoxEtAl2026}. SPX contracts are cash-settled, European-style index options with both AM- and PM-settled expirations. We focus the reported tables on closing-price-settled (PM-settled) contracts to avoid mixing monthly opening-price-settled (AM-settled) index options with otherwise similar PM-settled weekly contracts.

The option preprocessing uses bid and ask prices, midpoints, implied volatilities recovered from midpoint inversion when needed, forward prices, discount factors, open interest, and volume. We require positive bid/ask quotes, screen out quotes with relative bid-ask spreads above 0.60, restrict log-moneyness to $[-0.35,0.35]$ for SPX, and use out-of-the-money (OTM) quotes plus near-at-the-money (ATM) representatives. Deep in-the-money (ITM) calls and puts carry large intrinsic values and can dominate dollar errors while contributing little new information about the event premium. The OTM plus ATM protocol retains downside, upside, and straddle information while reducing redundant put-call replication.

After the settlement restriction, duplicate contract observations and estimation/evaluation contract overlap are both zero. This matters because the held-out evaluation design is intended to evaluate unseen event-spanning contracts, not differences between otherwise similar AM and PM settlement contracts.

\begin{figure*}[htp]
    \centering
    \includegraphics[width=\textwidth]{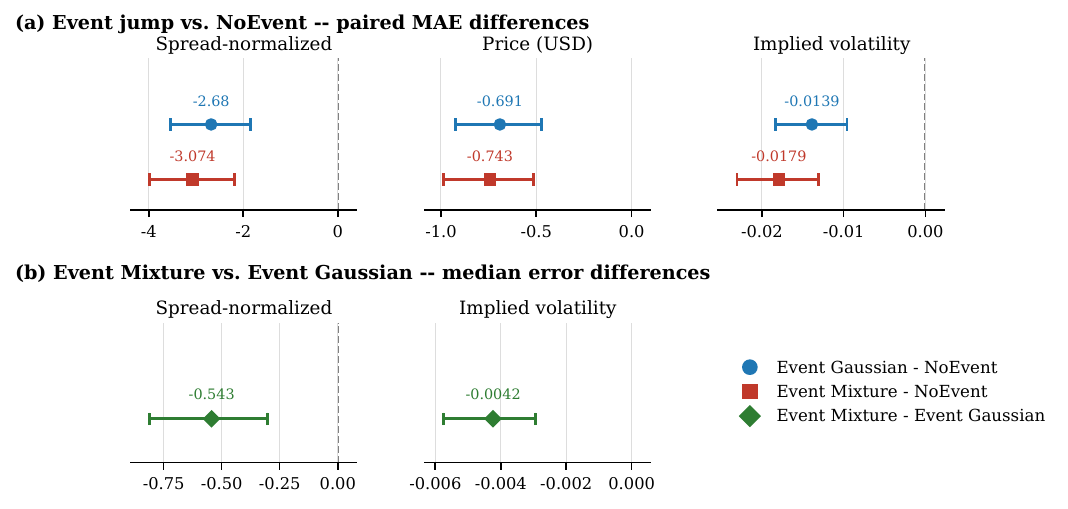}
    \caption{Event-level paired bootstrap differences in SPX specification with continuous IV cap $c_{\max}=1.00$. Points are paired differences and bars are 95\% confidence intervals; negative values favor the first model. Panel (a) reports mean-MAE differences relative to \NoEvent, while panel (b) reports direct \EMix-\EGauss median-error differences.}
    \Description{Five-panel forest plot of event-level paired bootstrap differences. Panel (a) shows spread-normalized, price, and implied-volatility mean absolute error differences for Gaussian and mixture models relative to no-event. Panel (b) shows median spread-normalized and median implied-volatility error differences for mixture relative to Gaussian. All intervals lie to the left of zero.}
    \label{fig:bootstrap}
\end{figure*}
\section{Results}
\label{sec:results}
\subsection{Main SPX Pricing Validation}

Table~\ref{tab:main_spx} reports the main SPX held-out result. Throughout the result tables, Price denotes price MAE, Spread denotes spread-normalized MAE, Med. Spr. denotes median spread-normalized absolute error, BA denotes bid-ask containment, IV denotes IV MAE, and Med. IV denotes median IV error. Bid-ask containment is strict: a predicted price must fall inside the quoted bid-ask interval, so short-dated tight spreads make relative lifts more informative than levels. A single Gaussian jump captures most of the improvement, while \EMix lowers IV MAE from 0.1080 to 0.0926 and delivers the best robust median errors. The neural benchmark rows are included for comparability and interpreted in Section~\ref{sec:neural_benchmark}.

\begin{table}[htp]
\centering
\caption{Held-out pricing results across parametric and neural scheduled-jump specifications. The continuous IV cap is $c_{\max}=1.00$. Lower is better except bid-ask containment. Short labels denote \NoEvent, \EGauss, \EMix, \ENeural, and \ENeuralCal.}
\label{tab:main_spx}
\small
\setlength{\tabcolsep}{2.0pt}
\begin{tabular*}{\columnwidth}{@{\extracolsep{\fill}}lrrrrrr@{}}
\toprule
Model & Price & Spread & Med. Spr. & BA & IV & Med. IV \\
\midrule
\textsc{NoEvent} & 7.917 & 69.231 & 14.008 & 1.58\% & 0.1080 & 0.0561 \\
\textsc{Gaussian} & 7.299 & 66.918 & 9.417 & 3.45\% & 0.0970 & 0.0370 \\
\textsc{Mixture2} & \textbf{7.255} & \textbf{66.601} & \textbf{8.915} & \textbf{4.62\%} & \textbf{0.0926} & \textbf{0.0305} \\
\textsc{Neural} & 8.122 & 70.253 & 13.634 & 2.22\% & 0.1054 & 0.0517 \\
\textsc{Neural-Cal} & 7.295 & 66.893 & 9.434 & 3.56\% & 0.0971 & 0.0371 \\
\bottomrule
\end{tabular*}
\end{table}

Because short-dated dollar prices vary sharply across moneyness, median spread-normalized and median IV errors provide a complementary view of the typical held-out contract. Event-jump models deliver their largest gains on these robust metrics.

Figure~\ref{fig:bootstrap} shows that the paired bootstrap intervals lie below zero for \EGauss and \EMix relative to \NoEvent; the bottom panels separately isolate \EMix's incremental value over \EGauss on median errors.

\subsection{SSVI Surface-Family Robustness and Contaminated-Surface Stress Test}

Table~\ref{tab:robustness_checks} Panel A repeats the pricing test with a compact SSVI-family continuous surface. Cross-family error levels are not the object of this check because SSVI and the polynomial surface emphasize different error criteria. The within-family comparison rules out a polynomial-surface artifact: adding a scheduled Gaussian event component improves the SSVI surface across price, spread-normalized, and implied-volatility metrics.

\begin{table}[t]
\centering
\caption{SSVI surface-family robustness and contaminated-surface stress test. Panel A uses a compact SSVI-family continuous surface. Panel B fits the no-event surface on both non-spanning contracts and event-spanning training quotes. Lower is better except bid-ask containment.}
\label{tab:robustness_checks}
\small
\setlength{\tabcolsep}{1.8pt}
\textit{Panel A: SSVI surface-family robustness check.}\\[-0.4ex]
\begin{tabular*}{\columnwidth}{@{\extracolsep{\fill}}lrrrrrr@{}}
\toprule
Model & Price & Spread & Med. Spr. & BA & IV & Med. IV \\
\midrule
\NoEventSSVI & 10.550 & 52.432 & 14.696 & 1.30\% & 0.1128 & 0.0834 \\
\EGaussSSVI & \textbf{9.995} & \textbf{50.433} & \textbf{12.254} & \textbf{1.79\%} & \textbf{0.1019} & \textbf{0.0686} \\
\bottomrule
\end{tabular*}
\par\vspace{0.8ex}
\textit{Panel B: Contaminated-surface stress test.}\\[-0.4ex]
\begin{tabular*}{\columnwidth}{@{\extracolsep{\fill}}llrrr@{}}
\toprule
Spec. & Surface fit set & Price & Spread & IV \\
\midrule
\textsc{NoEvent} & Non-spanning & 7.917 & 69.231 & 0.1080 \\
\NoEventContam & Non-spanning + event training & 0.380 & 2.962 & 0.0283 \\
\bottomrule
\end{tabular*}
\end{table}

Panel B gives the diagnostic failure mode: once event-spanning training quotes enter the no-event surface fit, held-out price MAE falls to 0.380, far below the best valid specification's 7.255. This performance reflects absorbed event premia and local interpolation, not scheduled-jump identification.

\subsection{Event Types, Jump Estimates, and Option-Combination Pricing Tests}
Table~\ref{tab:eventtype_jumps} shows that scheduled-jump pricing is strongest for CPI and FOMC and weaker for NFP, where the fitted jump scale often hits the lower bound. Comparable quote counts, relative spreads, and near-ATM call/put availability suggest that the NFP result is not a simple liquidity artifact.

\begin{table}[htp]
\centering
\caption{Event-type heterogeneity and quote quality. Panel A reports \EMix improvement over \NoEvent and fitted jump summaries. Event counts are 39/23/39 for CPI/FOMC/NFP. Panel B reports median quote and liquidity measures. BA lift is the percentage-point increase in bid-ask containment over \NoEvent; LB denotes lower-bound share: 0.10\% jump scale for \EMix and $\gamma_e=0$ for \ENeuralCal. Rel. spread is median relative bid-ask spread; ATM pair is the median near-ATM call/put-pair availability indicator.}
\label{tab:eventtype_jumps}
\small
\setlength{\tabcolsep}{1.4pt}
\textit{Panel A: Pricing improvements and fitted jump summaries.}\\[-0.4ex]
\begin{tabular*}{\columnwidth}{@{\extracolsep{\fill}}lrrrrrr@{}}
\toprule
Event & Med. Spread $\Delta$ & IV MAE $\Delta$ & BA lift & Jump std. & EMix LB & $\gamma$ LB \\
\midrule
CPI  & 5.793 & 0.0205 & 3.66\% & 0.91\% & 20.5\% & 20.5\% \\
FOMC & 7.197 & 0.0150 & 4.34\% & 0.81\% & 30.4\% & 30.4\% \\
NFP  & 2.903 & 0.0094 & 1.49\% & 0.10\% & 66.7\% & 66.7\% \\
\bottomrule
\end{tabular*}
\par\vspace{0.8ex}
\textit{Panel B: Quote and liquidity measures.}\\[-0.4ex]
\setlength{\tabcolsep}{2.2pt}
\begin{tabular*}{\columnwidth}{@{\extracolsep{\fill}}lrrrrr@{}}
\toprule
Event & Total quotes & Spanning & Held-out & Rel. spread & ATM pair \\
\midrule
CPI  & 564 & 506 & 126 & 0.087 & 1.00 \\
FOMC & 503 & 437 & 109 & 0.085 & 1.00 \\
NFP  & 496 & 423 & 106 & 0.087 & 1.00 \\
\bottomrule
\end{tabular*}
\end{table}

\begin{figure}[htp]
    \centering
    \includegraphics[width=\columnwidth]{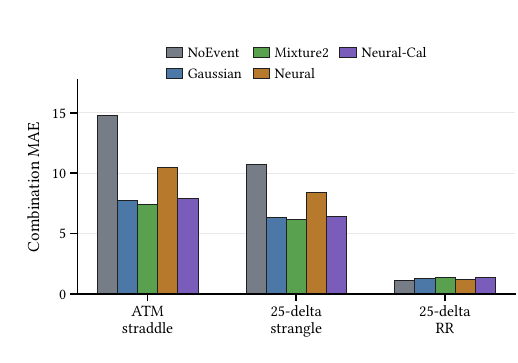}
    \caption{Option-combination pricing errors in the main SPX specification. Event-jump models improve ATM straddles and 25-delta strangles but not 25-delta risk reversals.}
    \Description{Grouped bar chart of option-combination MAE for ATM straddle, 25-delta
    strangle, and 25-delta risk reversal. Bars compare no-event, Gaussian event,
    mixture event, pure neural, and calibrated neural models.}
    \label{fig:portfolio_tests}
\end{figure}

Figure~\ref{fig:portfolio_tests} localizes the pricing gains to event-volatility combinations. ATM straddles and 25-delta strangles are direction-agnostic volatility exposures, whereas risk reversals load on directional skew. The absence of risk-reversal gains shows that the priced scheduled jump is mainly variance/convexity exposure, even though \EMix can represent asymmetric event beliefs; we use it as a pricing family, not as a label for realized macro regimes.

\subsection{Amortized Neural Benchmark and Event-Specific Scale}
\label{sec:neural_benchmark}

The final two rows of Table~\ref{tab:main_spx} report the neural scheduled-jump benchmarks. Pure cross-event \ENeural lowers IV MAE from 0.1080 to 0.1054 but does not improve mean spread-normalized or mean dollar errors. Event-level paired bootstrap against \NoEvent (Table~\ref{tab:neural_bootstrap}) gives an IV difference of -0.0048 with 95\% confidence interval [-0.0073,-0.0021], while the spread-error confidence interval [-0.330,1.613] crosses zero. The level differences in Table~\ref{tab:main_spx} are pooled quote-level MAEs, whereas Table~\ref{tab:neural_bootstrap} reports paired event-level contrasts, so the magnitudes need not coincide. This identifies a limited transfer channel: the non-spanning pre-event state contains IV-level event-risk information, but target-event option quotes remain necessary to recover jump scale.

\ENeuralCal sharpens this conclusion. Holding the leave-one-event-out neural mixture shape fixed and fitting a single target-event scalar $\gamma_e$ restores performance to the \EGauss level: price MAE is 7.295 versus 7.299 for \EGauss, spread-normalized MAE is 66.893 versus 66.918, and IV MAE is 0.0971 versus 0.0970. The bootstrap IV difference between \ENeuralCal and \EGauss is 0.00014 with confidence interval [0.00002,0.00034]. Thus \ENeuralCal is statistically distinguishable from \EGauss but economically close.

\ENeuralCal remains significantly worse than \EMix on every metric (IV difference 0.0054, confidence interval [0.0032,0.0078]). The $\gamma_e$ scalar reduces IV error by 0.0092 relative to pure \ENeural, about 1.9 times the pure cross-event shape-transfer improvement of 0.0048.

\begin{table}[htp]
\centering
\caption{Neural benchmark paired bootstrap differences. Each cell reports difference [95\% confidence interval]; negative means lower error for the first model. Columns report spread-normalized MAE, price MAE, and IV MAE differences. Short labels: N = \ENeural, C = \ENeuralCal, P = \NoEvent, G = \EGauss, M = \EMix.}
\label{tab:neural_bootstrap}
\footnotesize
\setlength{\tabcolsep}{1.5pt}
\begin{tabular*}{\columnwidth}{@{\extracolsep{\fill}}lccc@{}}
\toprule
Comparison & Spread & Price & IV \\
\midrule
N--P & 0.578 [-0.330,1.613] & 0.063 [-0.163,0.330] & -0.0048 [-0.0073,-0.0021] \\
C--P & -2.706 [-3.575,-1.875] & -0.695 [-0.927,-0.477] & -0.0138 [-0.0182,-0.0096] \\
C--N & -3.284 [-4.302,-2.421] & -0.758 [-1.039,-0.522] & -0.0092 [-0.0125,-0.0061] \\
C--G & -0.026 [-0.039,-0.013] & -0.004 [-0.006,-0.002] & 0.00014 [0.00002,0.00034] \\
C--M & 0.368 [0.224,0.527] & 0.048 [0.028,0.072] & 0.0054 [0.0032,0.0078] \\
\bottomrule
\end{tabular*}
\end{table}

\begin{figure}[htp]
    \centering
    \includegraphics[width=\columnwidth]{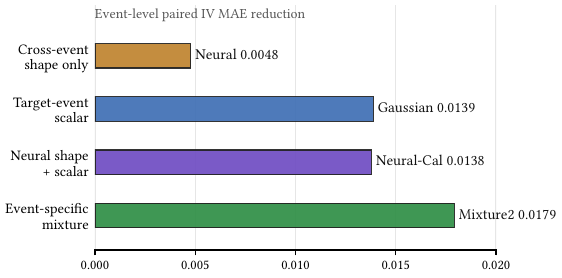}
    \caption{Scale-shape attribution for scheduled event-jump pricing. Bars report event-level paired IV MAE reductions relative to \NoEvent; they are nested benchmark differences, not an exact additive decomposition.}
    \Description{Horizontal bar chart of event-level paired IV MAE reductions relative to the no-event baseline. Bars compare cross-event neural shape only, target-event scalar calibration, neural shape plus scalar calibration, and event-specific mixture shape.}
    \label{fig:shape_scale}
\end{figure}

Figure~\ref{fig:shape_scale} summarizes the nested benchmark attribution.

The fitted $\gamma_e$ lower-bound share is 40.6\%, directly quantifying a scale-identification limit: for these event-level fits, the target-event training quotes do not support retaining the transferred neural jump scale, so the calibrated variant sets the scheduled-jump scale to its lower bound. By event type, the lower-bound shares are CPI 20.5\%, FOMC 30.4\%, and NFP 66.7\%, exactly matching the parametric mixture's lower-bound shares in Table~\ref{tab:eventtype_jumps}. This pattern shows that weak NFP scale identification is driven by the data rather than cross-event amortization.
Increasing neural training from 40 to 200 epochs leaves IV-level metrics essentially unchanged and leaves \ENeuralCal nearly identical. The binding constraint is therefore informational rather than optimization-driven.

\section{Conclusion and Future Work}
\label{sec:conclusion}

Scheduled macro announcements create known-time risk that a flexible no-event implied-volatility surface can obscure. The main point is methodological: event-spanning fit is not event-risk identification. As the contaminated-surface stress test shows, a surface fitted to event-spanning quotes can absorb event premia rather than identify the scheduled jump. When the continuous surface is flexible, identification requires fixing it on non-spanning expiries before using event-spanning training contracts to fit the scheduled jump and held-out event-spanning contracts to test it.

Within this information set, the evidence points to a scale-shape attribution of scheduled event risk. Gaussian jumps capture most of the held-out pricing gain, event-specific mixtures add residual higher-moment information, and the neural MDN benchmark shows that pure cross-event shape transfer is limited. A single target-event scale parameter restores Gaussian-level performance, but the calibrated model remains below event-specific mixture calibration. The matching lower-bound shares from the parametric mixture and calibrated neural scale suggest that the CPI/FOMC versus NFP boundary reflects the same scale-identification limit rather than a model artifact. The object identified here is scheduled-event variance and convexity exposure rather than directional event skew. Shape partly transfers across announcements, but jump scale must come from target-event quotes.

Future work can test this boundary in intraday, cross-market, and richer event
panels, where target-event scale may become identifiable earlier and
cross-event jump-shape transfer may be more stable.

\balance
\bibliographystyle{ACM-Reference-Format}
\bibliography{references}

\end{document}